\documentclass{revtex4}
\usepackage{psfig}
\usepackage{subfigure}
\usepackage{graphicx}
\usepackage{doublespace}
\usepackage{amsmath}

\begin{document}

\author{Francesco Sorrentino${}^{1, 2}$, Gilad Barlev${}^{2}$,  Adam B. Cohen${}^{1}$,  Edward Ott${}^{1}$}
\affiliation{${}^1$ Institute for Research in Electronics and Applied Physics, University of Maryland, College Park, Maryland 20742, USA. \\ ${}^2$ Universit{\`a} degli Studi di Napoli Parthenope, 80143 Napoli, Italy. }

\begin{abstract}
In past works, various schemes for adaptive synchronization of chaotic systems have been proposed. The stability of such schemes is central to their utilization. As an example addressing this issue, we consider a recently proposed adaptive scheme for maintaining
the synchronized state of identical coupled chaotic systems in the presence of {a priori} unknown slow temporal drift in the couplings.  For this illustrative example, we develop an extension of the master stability function technique to study synchronization stability with adaptive coupling. Using this formulation, we examine local stability of synchronization for typical chaotic orbits and for unstable periodic orbits within the synchronized chaotic attractor (bubbling). Numerical experiments illustrating the results are presented.
We observe that the stable range of synchronism can be sensitively dependent on the adaption parameters, and we discuss the strong implication of bubbling for practically achievable adaptive synchronization.
\end{abstract}

\title{The stability of adaptive synchronization of  chaotic systems}

\maketitle

\textbf{We consider an adaptive scheme for maintaining
the synchronized state in a network of identical coupled chaotic systems in the presence of {a priori} unknown slow temporal drift in the couplings.
Stability of this scheme is addressed through an extension of 
the master stability function technique to include adaptation. We observe that noise and/or slight nonidenticality between the coupled systems can be responsible for the occurrence of intermittent bursts of large desynchronization events  ({bubbling}).
Moreover, our numerical computations show that, for our adaptive synchronization scheme, the parameter space region corresponding to bubbling can be rather substantial. This observation becomes important to experimental realizations of adaptive synchronization, in which small mismatches in the parameters and noise cannot be avoided. We also find that, for our coupled systems with adaptation, bubbling can be caused by a slow drift in the coupling strength.}

\section{Introduction}
It has been shown \cite{FujiYama83,Afraim,Pe:Ca} that, in spite of their random-like behavior, the states $x_i(t)$ $(i=1,2,...,N)$ of a collection of $N$ interacting chaotic systems that are identical can synchronize (i.e., be attracted toward a common chaotic evolution, $x_1(t)=x_2(t)=...=x_N(t)$) provided that they are properly coupled. This phenomenon has been the basis for proposals for secure communication \cite{Cu:Op,Argyris,Feki}, system identification \cite{Abarbanel, Abarbanel2,Abarbanel4,IDTOUT}, data assimilation \cite{So:Ott:Day,Duane}, sensors \cite{SOTT2}, 
information encoding and transmission \cite{Ha:Gr:Ott,Dr:He:An:Ott}, multiplexing \cite{Ts:Su}, combatting channel distortion \cite{Sh:Ott}, etc. In all of these applications it is typically assumed that one has accurate knowledge of the interaction between the systems, allowing one to choose the appropriate coupling protocol at each node (here we use the network terminology, referring to the $N$ chaotic systems as $N$ nodes of a connected network whose links $(i,j)$ correspond to the input that node $i$ receives from node $j$).
In a recent paper \cite{SOTT},  an adaptive strategy was proposed for maintaining synchronization between identical coupled chaotic dynamical systems in the presence of a priori unknown, slowly time varying coupling strengths (e.g., as might arise from temporal drift of environmental parameters). This strategy was successfully tested on computer simulated networks of many coupled dynamical systems in which, at each time, every node receives only one aggregate signal representing the superposition of signals transmitted to it from the other network nodes. In addition, the strategy has also been successfully implemented in an experiment on coupled optoelectronic feedback loops \cite{EXP}. Furthermore, a more generalized adaptive strategy, suitable for sensor applications, has also been proposed \cite{SOTT2}.

In past works, various other schemes for adaptive synchronization of chaos 
have also been proposed \cite{rr1,rr2,rr3,rr4,rr5,rr6,rr7,rr8,Zh:Ku06,Delellis2}.
So far, in all these studies, when the question of stability of the considered adaptive schemes has been studied, the question has been addressed using the Lyapunov function method (see e.g., \cite{rr3,rr4,rr6,rr8}), which provides a sufficient but not necessary condition for stability. While this technique has the advantage that it can sometime yield global stability conditions, it also has the disadvantages that its applicability is limited to special cases, and its implementation, when possible, requires nontrivial system specific analysis.
In this paper, we address the stability of adaptive synchronization for the example of the scheme discussed in Ref. \cite{SOTT}. 
In particular, our analysis will extend the previously developed stability analysis of chaos synchronization by the master stability function technique \cite{FujiYama83,Pe:Ca} to include adaptation.  
We will observe that the range in which the network eigenvalues are associated with stability,  is dependent on the choice of the parameters of the adaptive strategy.
The type of analysis we present, while for a specific illustrative adaptive scheme, can be readily applied to other adaptive schemes  (e.g., those in \cite{Zh:Ku06,Delellis2}).

As compared to the Lyapunov technique, master stability techniques are much more generally applicable  but they provide conditions for local, rather than global stability. We also note that, within that context, the master stability technique allows one to distinguish between stability of typical chaotic orbits and stability of atypical orbits within the synchronizing chaotic attractor [i.e., stability to `bubbling' \cite{Ash1, Ash2,bub1,bub2,restr_bubbl,OB}; see Secs. III and IV].

In Sec. II we review the adaptive synchronization strategy formulation of Ref. \cite{SOTT}, which applies to a network of chaotic systems with unknown temporal drifts of the couplings. In Sec. III, we present a master stability function approach to study linear stability of the synchronized solution in the presence of adaptation; we also consider a generalized formulation of our adaptive strategy and study its stability. Numerical simulations are finally presented in Sec. IV. Our work in Sec. IV highlights the important effect of bubbling in the dynamics.

\section{Adaptive strategy formulation}

As our example of the application of the master stability technique to an adaptive scheme, we consider the particular scheme presented in Ref. \cite{SOTT}.
 To provide background, in this section we present a brief exposition of a formulation similar to that in Ref. \cite{SOTT}, as motivated by the situation where the couplings are unknown and drift with time. We consider a situation where the dynamics at each of the network nodes is described by,
\begin{equation}
\begin{split}
\dot{x}_i(t)=F(x_i(t))+ \gamma \Gamma [\sigma_i(t) r_i(t)-H(x_i(t))], \qquad i=1,...,N, \label{net}
\end{split}
\end{equation}
 where, $x_i$  is the $m$-dimensional state of system $i=1,...,N$;  $F(x)$ determines the dynamics of an uncoupled ($\gamma \rightarrow 0$) system (hereafter assumed chaotic), $F:R^m \rightarrow R^m$; $H(x)$ is a scalar output function, $H:R^m \rightarrow R$. We take $\Gamma$ to be a constant $m$-vector, $\Gamma=[\Gamma_1,\Gamma_2,...,\Gamma_m]^T$, with $\sum_i \Gamma_i^2=1$, and the scalar $\gamma$ is a constant characterizing the strength of the coupling. The  scalar signal each node $i$ receives from the other nodes in the network is,
\begin{equation}
r_i(t)=\sum_j A_{ij}(t) H(x_j(t)). \label{ri}
\end{equation}
 The quantity $A_{ij}(t)$ is an adjacency matrix whose value specifies the strength of the coupling from node $j$ to node $i$. We note that if
 \begin{equation}
 \sigma_i(t)=[\sum_j A_{ij}]^{-1} \label{goal}
 \end{equation}
 then Eq. (\ref{net}) admits a synchronized solution,
 \begin{equation}
 x_1(t)=x_2(t)=...=x_N(t)=x_s(t), \label{xs}
 \end{equation}
  where $x_s(t)$ satisfies
  \begin{equation}
  \dot{x}_s(t)=F(x_s(t)), \label{five}
  \end{equation}
 which corresponds to the dynamics of an isolated system.
 We regard the $A_{ij}(t)$ as unknown at each node $i$, while the only external information available at node $i$ is its received signal (\ref{ri}). The goal of the adaptive strategy is to adjust $\sigma_i(t)$ so as to maintain synchronism in the presence of slow, a priori unknown time variations of the quantities $A_{ij}(t)$. That is, we wish to maintain approximate satisfaction of Eq. (\ref{goal}). For this purpose, as discussed in Ref. \cite{SOTT}, our scheme can be extended to the case where the output function is $\ell$-dimensional, $H:R^m \rightarrow R^{\ell}$, where $\ell<m$ and $\Gamma$ is an $\ell \times m$ dimensional matrix. For simplicity we consider $\ell=1$.
 We assume that each node independently implements an adaptive strategy. At each system node $i$, we define the exponentially weighted synchronization error $\psi_i=<(\sigma_i r_i -H(x_i))^2>_{\nu}$, where
 \begin{equation}
 <G(t)>_{\nu}=\int^t G(t') e^{-\nu (t'-t)} dt',
 \end{equation}
  and we evolve $\sigma_i(t)$ so as to minimize this error (a slightly more general approach is taken in \cite{SOTT}).
 Hence we set ${\partial \psi_i}/{\partial \sigma_i}$ equal to zero to obtain,
\begin{equation}
\sigma_i(t)=\frac{<H(x_i(t)) r_i(t)>_{\nu}}{<r_i(t)^2>_{\nu}}=\frac{p_i(t)}{q_i(t)}. \label{sigma}
\end{equation}
By virtue of ${d<G(t)>_{\nu}}/{dt}=-\nu <G(t)>_{\nu}+G(t)$, we obtain the numerator and the denominator on the right hand side of Eq. (\ref{sigma}) by solving the differential equations,
\begin{subequations} \label{quattro}
\begin{align}
\dot{p}_i(t)=-\nu p_i(t)+ r_i(t) H(x_i(t)), \label{quattroa}\\
\dot{q}_i(t)=-\nu q_i(t)+ r_i(t)^2. \label{quattrob}
\end{align}
\end{subequations}
Since the dynamics of $A_{ij}(t)$ is imagined to occur on a timescale which is slow compared to the other dynamics in the network, we can approximate $A_{ij}(t)$ as constant $A_{ij}$. This essentially assumes that we are dealing with perturbations from synchronization whose growth rates (in the case of unstable synchronization) or damping rates (in the case of stable synchronization) have magnitudes that substantially exceed $|A_{ij}^{-1}(t) d A_{ij}/dt|$. Under this assumption, we note that  Eqs. (\ref{net}), (\ref{sigma}), and (\ref{quattro}) admit a synchronized solution, given by Eqs. (\ref{xs}), (\ref{five}), and
\begin{subequations}\label{cinque}
\begin{align}
\dot{p}_i^s=-\nu p_i^s +(\sum_j A_{ij}) H(x^s)^2, \quad i=1,...,N,  \label{cinqueb}\\
\dot{q}_i^s=-\nu q_i^s+(\sum_j A_{ij})^2 H(x^s)^2, \quad i=1,...,N. \label{cinquec}
\end{align}
\end{subequations}
To simplify the notation, in what follows,  we take $DF^s(t)=DF(x^s(t))$, $H^s(t)=H(x^s(t))$, and $DH^s(t)=DH(x^s(t))$; 
e.g., we can now write,
\begin{equation}
\begin{split}
p_i^s=k_i <(H^s)^2>_{\nu}, \\
q_i^s=k_i^2 <(H^s)^2>_{\nu}, \label{sei}
\end{split}
\end{equation}
where $k_i=(\sum_j A_{ij})$. If the synchronization scheme is locally stable, we expect that the synchronized solution (\ref{xs}),(\ref{five}), and (\ref{cinque}) will be maintained under slow time evolution of the couplings $A_{ij}(t)$.

\section{Stability analysis}

\subsection{Linearization and master stability function}

Our goal is to study the stability of the reference solution (\ref{xs}),(\ref{five}), and (\ref{cinque}).
By linearizing Eqs. (\ref{net}) and (\ref{quattro}) about (\ref{five}), and (\ref{cinque}), we obtain,

\begin{subequations}\label{d}
\begin{align}
\delta \dot{x}_i=DF^s \delta x_i+ \gamma \Gamma \biggl\{ DH^s  \biggl[k_i^{-1} \sum_j A_{ij} \delta x_j - \delta x_i \biggl]+ \frac{H^s}{k_i^2 <(H^s)^2>_{\nu}}  \epsilon_i \biggl\}, \quad i=1,...,N, \label{dxi}\\
 \dot{\epsilon}_i=-\nu \epsilon_i -H^s DH^s  k_i \biggl[\sum_j A_{ij} \delta x_j- k_i \delta x_i \biggl], \quad i=1,...,N, \label{deps}
\end{align}
\end{subequations}
where we have introduced the new variable $\epsilon_i(t)=k_i \delta p_i(t)- \delta q_i(t)$.

Equations (\ref{d}) constitute a system of $(m+1)N$ coupled equations. In order to simplify the analysis, we seek to decouple this system into $N$ independent systems, each of dimension $(m+1)$. For this purpose we seek a solution where $\delta x_i$ is in the form $\delta x_i=c_i \bar{x}(t)$, where $c_i$ is a time independent scalar that depends on $i$ and $\bar{x}(t)$ is a $m$-vector that depends on time but not on $i$. Substituting in Eqs. (\ref{dxi}),(\ref{deps}), we obtain,
\begin{subequations}\label{ciPS}
\begin{align}
\dot{\bar{x}}=DF^s  \bar{x}+ \gamma \Gamma \biggl[  \frac{\sum_j A_{ij}  c_j }{k_i c_i} - 1 \biggl] DH^s \bar{x}+ \frac{ \gamma \Gamma H^s}{ c_i k_i^2 <(H^s)^2>_{\nu}}  \epsilon_i, \quad i=1,...,N,  \label{ciPE2} \\
 \dot{\epsilon}_i=-\nu \epsilon_i -   k_i \biggl[\sum_j A_{ij} c_j - k_i c_i \biggl] H^s DH^s \bar{x}, \quad i=1,...,N.\label{ciSE}
\end{align}
\end{subequations}
 To make  Eqs. (\ref{ciPS}) independent of $i$, we  consider $\beta(t)=\epsilon_i(t)/[c_i {k_i}^2 (\alpha-1)]$ and ${\sum_j A_{ij}  c_j }= \alpha {k_i c_i}$, where $\alpha$ is a quantity independent of $i$. Namely, the possible values of $\alpha$ are the eigenvalues, $ \bf{A'}  \bf{c} = \alpha {\bf{c}}$, corresponding to  linearly independent eigenvectors ${\bf{c}}=[c_1,c_2,...,c_N]^T$, where ${\bf{A'}}=\{A'_{ij}\}=\{{k_i}^{-1} A_{ij}\}$. This gives,
\begin{subequations}\label{IND}
\begin{align}
 \dot{\bar{x}}=DF^s  \bar{x}- \gamma (1-\alpha) \biggl[ \Gamma  DH^s \bar{x} +  \Gamma \frac{H^s \beta}{ <(H^s)^2>_{\nu}} \biggl], \label{INDa} \\
 \dot{\beta}=-\nu \beta -   H^s DH^s \bar{x},
\end{align}
\end{subequations}
 which is independent of $i$, but depends on the eigenvalue $\alpha$. Considering the typical case where there are $N$ distinct eigenvalues of the $N \times N$ matrix ${\bf{A'}}$, we see that Eqs. (\ref{ciPS}) constitute $N$ decoupled linear ordinary differential equations for the synchronization perturbation variables $\bar{x}$ and $\beta$.  All the rows of  $ \bf{A'}$ sum to $1$. Therefore $ \bf{A'}$ has at least one eigenvalue $\alpha=1$, corresponding to the eigenvector $c_1=c_2=...=c_N=1$. Furthermore, since $A'_{ij} \geq 0$ for all $(i,j)$, we have by the Perron-Frobenius theorem that $\alpha \leq 1$, and thus $(1-\alpha) \geq 0$. For $\alpha=1$, Eq. (\ref{INDa}) becomes,
 \begin{equation}
 \dot{\bar{x}}=DF^s  \bar{x}.
\end{equation}
This equation reflects the chaos of the reference synchronized state (Eq. (\ref{five})) and (because all the $c_i$ are equal) is associated with perturbations which are tangent to the synchronization manifold and are therefore irrelevant in determining synchronization stability. Stability of the synchronized state thus demands that Eqs. (\ref{ciPS}) yield exponential decay of $\bar{x}$ and $\beta$ for all the $(N-1)$ eigenvalues $\alpha$, excluding this $\alpha=1$ eigenvalue.

Then it becomes possible to introduce a master stability function \cite{FujiYama83,Pe:Ca}, $M(\xi)$, that associates  the maximum Lyapunov exponent of system (\ref{IND}) with $\xi=\gamma(1-\alpha)$. In so doing, one decouples the effects of the network topology (reflected in the eigenvalues $\alpha$ and hence the relevant values of $\xi=\gamma(1-\alpha)$) from the choices of $F,H,\nu$. In general an eigenvalue, and hence also $\xi$, can be complex. For simplicity, in our discussion and numerical examples to follow, we assume that the eigenvalues are real (which is for instance the case when the adjacency matrix is symmetric). For any given value of $\gamma$ stability demands that $M(\xi)<0$ for all those values of $\xi=\gamma(1-\alpha)$ corresponding to the eigenvalues $\alpha \neq 1$.

Following Refs. \cite{Ba:Pe02,Ni:Mo,Bocc1,Bocc2}, we now introduce the following definition of \emph{synchronizability}. Let us assume that the master stability function $M(\xi)$ is negative in a bounded interval of values of $\xi$, say $[\xi^-,\xi^+]$. Then, in order for the network to synchronize, two conditions need to be satisfied, (i) $\xi^-<\gamma(1-\alpha_{min})$, and (ii) $\xi^+>\gamma(1-\alpha_{max})$, where $\alpha_{min}$ ($\alpha_{max}$) is the smallest (largest) network eigenvalue over all the eigenvalues $\alpha \neq 1$. The network synchronizability is defined as the width of the range of values of $\gamma$, for which $M(\xi)<0$. 
 Assuming that $\alpha_{min}$ and $\alpha_{max}$  are assigned  (e.g., the network topology is given),  then the network synchronizability increases with the ratio $\xi^+/\xi^-$. In what follows, we will compare different adaptive strategies in terms of their effects on the  synchronizability ratio $\xi^+/\xi^-$. 

In our analysis above, since we divide by $k_i$, we have implicitly assumed that all the $k_i \neq 0$, i.e., that every node has an input. There is, however, a case of interest where this is not so, and this case requires separate consideration. In particular, say there is one and only one special node (which we refer to as the maestro or sender) that has no inputs, but sends its output to other nodes (which interact with each other), and we give this special node the label $i=N$. Since node $N$ receives no inputs, we do not include adaption on this node, and we replace Eq.(1) for $i=N$ by $\dot{x}_N(t)=F(x_N(t))$. In addition, when investigating the stability of the synchronized state, it suffices to set $\delta x_N(t)=0$ (i.e., not to perturb the maestro). Following the steps of our previous stability analysis, we again obtain Eqs. (\ref{ciPS}) and (\ref{IND}), but with important differences. Namely, Eqs. (\ref{ciPS}) now apply for $i=1,...,N-1$, the values of $\alpha$ in Eqs. (\ref{IND}) are now the eigenvalues of the $(N-1)\times(N-1)$ matrix $\{ A'_{ij} \}= \{ {k_i}^{-1} A_{ij} \}$ for $i,j=1,2,...,(N-1)$; i.e., only the interactions between the nodes $i,j \leq (N-1)$ are included in this matrix. Note that $k_i$ is still given by $\sum_{j=1}^N A_{ij}$, still including the input $A_{iN}$ from the maestro node. Also since $\delta x_N=0$, all of the eigenvalues represent transverse perturbations and are therefore relevant to stability. (This is in contrast to the case without a maestro in which we had to exclude an eigenvalue, i.e., $\alpha=1$ corresponding to $c_1=c_2=...=c_N=1$. For a similar discussion for the case of the standard master stability problem with no adaptation, see \cite{PC}.) The simplest case of this type (used in some of our subsequent numerical experiments) is the case $N=2$, where there is one receiver node ($i=1$) and one sender/maestro node ($i=2$). Since there is only one receiver node whose only input is received by the sender, $A$ reduces to the scalar $A=0$ and $\alpha=0$, yielding $\xi \equiv \gamma$.

As stated above, $x_s(t)$ in (\ref{xs}) is an orbit of the uncoupled system (\ref{xs}). In general, two types of orbits $x_s(t)$ are of interest: (i) a typical chaotic orbit on the relevant chaotic attractor of (\ref{xs}), and (ii) the orbit that is ergodic on the maximally synchronization-unstable invariant subset embedded within the relevant chaotic attractor of (\ref{xs}). Here, by `relevant chaotic attractor' we mean that, if the system (\ref{xs}) has more than one attractor, then we restrict attention to that attractor on which synchronized motion is of interest.   Also, in (i), by the word `typical', we mean orbits of (\ref{xs}) that ergodicly generate the measure that applies for Lebesgue almost every initial condition in the attractor's basin of attraction. In this sense, the orbit in (ii) is not typical. In general the criterion for stability as assessed by (ii) is more restrictive than that assessed by (i). Conditions in which the synchronized dynamics is stable according to (i), but unstable according to (ii), are referred to as the `bubbling' regime \cite{Ash1, Ash2,bub1,bub2,OB}. In previous work on synchronization of chaos \cite{Ash1, Ash2,bub1,bub2,restr_bubbl,OB}, it has been shown that, when the system is in the bubbling regime, small noise and/or small `mismatch' between the coupled systems can lead to rare, intermittent, large deviations from synchronism, called `desynchronization bursts' \footnote{In addition, to desynchronism bursts, it is also possible that a large desynchronization orbit event can result in capture of the system orbit on a desynchronized attractor with a so-called riddled basin of attraction (e.g., see Refs. \cite{Ash1,Ash2,bub1,Ott:Som,OB}). This possibility, although a typical occurrence in such situations, did not manifest itself in the particular system used for our numerical studies in Sec. IV. Thus we, henceforth, restrict our discussion to the case that bubbling is associated with bursts.}. By small system mismatch we mean that, for each node $i$, the functions $F$ in (\ref{net}) are actually different, $F \rightarrow F_i$, but that these differences are small (i.e., $|F_i(x) - F(x)|$ is small, where $F(x)$ now denotes a reference uncoupled system dynamics; e.g., $F_i$ averaged over $i$). With reference to our adaptive synchronization problem (\ref{net}), we shall see that, in addition to small noise and small mismatch in $F$, bursting can also be induced by slow drift in the unknown couplings $A_{ij}(t)$. From the practical, numerical perspective, the complete and rigorous application of the stability criterion (ii) is impossible, since there will typically be an infinite number of distinct invariant sets embedded in a chaotic attractor, and, to truly be sure of stability, each of these must be found and numerically tested. In practice, therefore, as done previously by others, we will evaluate stability for all the unstable periodic orbits embedded in the attractor up to some specified period. This will give a necessary condition for stability according to (ii), and furthermore, it has been argued and numerically verified in Ref. \cite{bub2} that stability, as assessed from a large collection of low period periodic orbits (and embedded unstable fixed points, if they exist in the relevant attractor), will extremely often yield the true delineation of the parameters of the bubbling regime, or, if not, an accurate approximation of it. Our numerical results of Sec. IV lend further support to this idea.

\subsection{Generalized adaptive strategy}

We now analyze a generalization of our adaptive strategy.  Namely, we replace Eq. (\ref{quattrob}) by,
\begin{equation}
\dot{p}_i(t)=-\nu p_i(t)+ [q_i(t)/p_i(t)] H(x_i(t))^2 Q\biggl( \frac{p_i(t) r_i(t)}{q_i(t) H(x_i(t))} \biggl),\label{quattrobis}
\end{equation}
where $Q(z)$ is an arbitrary function of $z$, normalized so that $Q(1) \equiv 1$. The key point is that at synchronism $\sigma_i r_i=H(x_i(t))$, corresponding to $p_i r_i=q_i H(x_i(t))$; and thus, since we take $Q(1)=1$, the synchronized solution is unchanged.
The stability analysis for this generalization is given in the Appendix I and results in the following master stability equations,
\begin{subequations}\label{INDA}
\begin{align}
 \dot{\bar{x}}=DF^s  \bar{x}- \xi \biggl[ \Gamma  DH^s \bar{x} +  \Gamma \frac{H^s \beta'}{ <(H^s)^2>_{\nu}} \biggl], \label{INDAa}  \\
 \dot{\beta'}=-\nu \beta' + (\phi-1) \frac{(H^s)^2}{<(H^s)^2>_{\nu}} \beta' +(\phi-2) H^s DH^s  \bar{x} \label{INDAb},
\end{align}
\end{subequations}
where $\phi={Q'}(1)$, and $Q'(1)$ denotes $dQ(z)/dz$ evaluated at $z=1$.
We then introduce a master stability function $M(\xi,\phi)$, that associates  the maximum Lyapunov exponent of system (\ref{INDA}) with $\xi=\gamma(1-\alpha)$ and $\phi$. 

Thus we expect that, when our modified adaptive scheme is stable, it will again relax to the desired synchronous solution. The difference between the stability of the modified scheme (Eqs. (\ref{IND})) and the stability of the original scheme (corresponding to Eqs. (\ref{INDA}) with $\phi=1$), is that, by allowing the freedom to choose the value of $\phi$, we can alter the stability properties of the synchronous state. We anticipate that, by properly adjusting $\phi$, we may be able to tailor the stability range to better suit a given situation.

In the case of $\phi=2$, Eq. (\ref{INDAb}) reduces to,
\begin{equation}
\dot{\beta'}=\big[\frac{(H^s)^2}{<(H^s)^2>_{\nu}} -\nu \big]  \beta', \label{ups0}
\end{equation}
which has a Lyapunov exponent $\lambda=\lambda_0-\nu$, where $\lambda_0$ is the time average of $(H^s)^2/<(H^s)^2>_{\nu}$, $\lambda_0\geq0$. For $\nu>\lambda_0$, Eq. (\ref{ups0}) implies that $\beta'$ decays to zero. Thus, if we choose a large enough value of $\nu$, stability of the synchronized state is determined by (\ref{INDAa}) with $\beta'$ set equal to zero, and Eq. (\ref{INDAa}) reduces to the master stability function for the determination of the stability of the system without adaptation \cite{Pe:Ca}. Therefore, in the case of $\phi=2$, $\nu> \lambda_0$, the stable range of $\gamma$ is independent of $\nu$ and is the same as that obtained for the case in which adaptation is not implemented ($\sigma\equiv 1$).

\section{Numerical experiments}

In our numerical experiments we consider the example of the following R\"ossler equation,
for which, $m=3$, $x(t)=(u(t),v(t),w(t))^T$,
\begin{equation}
F({{x}})=\left[\begin{array}{c}
    -v-w  \\
    u + a v \\
    b+ (u-c )  w
  \end{array} \right], \label{RC}
\end{equation}
with the parameters $a=b=0.2$, and $c=7$, and we use $H(x(t))=u(t)$, and $\Gamma=[1,0,0]^T$.
In Fig. 1 the master stability functions $M(\xi)$ calculated from Eq. (\ref{IND}) for the adaption scheme of Sec. II are plotted for three different values of $\nu$, i.e., $\nu=0.1,2,6$ (dashed, dashed/dotted, and dotted curves, respectively).
In addition, for comparison, we also plot the result of $M(\xi)$ computations for the case in which no adaptation is introduced, corresponding to the reduced system $\dot{\bar{x}}=[DF^s + \gamma (\alpha-1) \Gamma DH^s ] \bar{x}$ (solid curves). 
 The master stability function is shown in black (respectively, grey) for the cases that $x_s(t)$ is a typical chaotic orbit in the attractor (respectively, the maximally unstable periodic orbit embedded in the attractor for periodic orbits of period up to four surface of section piercings; see Appendix II for a brief account of how the unstable periodic orbits were obtained).
We say that synchronization is `high quality' stable in the range of $\xi$ for which $M(\xi)$ for all orbits (i.e., including the periodic orbits) is negative. As can be seen, by changing the parameter $\nu$, the $\xi$-range of stability  can be dramatically modified.  The bubbling range is given by the values of $\xi$ for which $M(\xi)<0$ for a typical orbit but $M(\xi)>0$ for the maximally unstable periodic orbit embedded in the attractor.
\begin{figure}[h]
\centering
\includegraphics[width=0.7\textwidth]{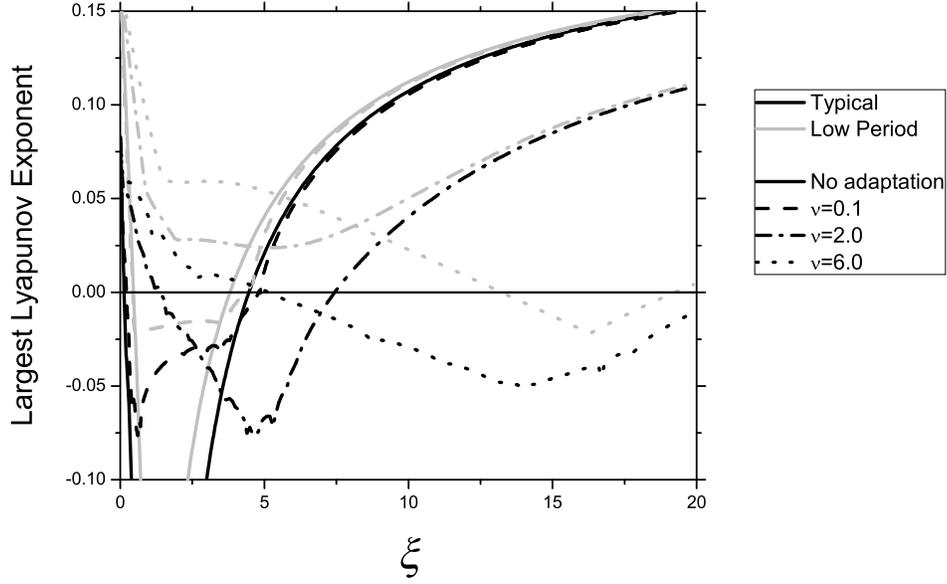}
\caption{The plot shows the master stability function $M(\xi)$ versus $\xi$ for for the case in which no adaptation was introduced, corresponding to $\sigma\equiv 1$ (black continuous line) and for three different values of $\nu$, i.e., $\nu=0.1,2,6$ (dashed and dotted lines). The master stability functions obtained by choosing $x_s(t)$ to be a typical chaotic orbit in the attractor (respectively, the maximally unstable periodic orbit embedded in the attractor of period up to four) are in black (respectively, grey). $F(x)$ is the R\"ossler equation (\ref{RC}), $H(x(t))=u(t)$, and $\Gamma=[1,0,0]^T$. \label{UNO}}
\end{figure}

Figure 2 is a $\xi-\nu$ level curve plot of the values assumed by the master stability function $M$ evaluated for $x_s(t)$ being a typical chaotic orbit. In the figure, the area of stability (corresponding to $M<0$) is delimited by the thick $0$-level contour line. From the figure, we see that the width of the range of stability increases with $\nu$.  In Figs. 3(a,b) a comparison between the areas of stability is given, for the cases in which $x_s(t)$ is a typical chaotic orbit in the attractor, and for the case that $x_s(t)$ is the maximally unstable periodic orbit embedded in the attractor of period up to four. The thick solid (respectively, dashed) curves bound the area in which the master stability function $M(\xi,\nu)$ is negative for $x_s(t)$ corresponding to a typical chaotic orbit in the R\"osller attractor (respectively, for $x_s(t)$ corresponding to the maximally unstable periodic orbit embedded in the attractor of period up to four). The bubbling area falls between the dashed and the continuous contour lines.

Interestingly, we see that for $1.2 \lesssim \nu \lesssim 3.2$, high-quality stability can never be achieved for any $\xi$, while, in contrast, stability with respect to typical chaotic orbits (i.e., with bubbling) is achievable. Let $\xi^+_t,\xi^-_t,\xi^+_p,\xi^-_p$ denote the upper $(+)$ and lower $(-)$ values of $\xi$ at the borders of the stability regions with respect to a typical (t) chaotic orbit and with respect to unstable periodic orbits (p) in the synchronizing attractor. E.g., high-quality synchronism applies for $\xi^+_p > \xi > \xi^-_p$ and the bubbling regime corresponds to $\xi^-_p > \xi > \xi^-_t$ or $\xi^+_t > \xi > \xi^+_p$. In terms of these quantities, useful measures for assessing the possibility of achieving stable synchronism for a given network topology are the `synchronizability' ratios \cite{Ba:Pe02,Ni:Mo,Bocc1,Bocc2},
\begin{equation}
s_t=\frac{\xi^+_t}{\xi^-_t}, \qquad s_p=\frac{\xi^+_p}{\xi^-_p}.
\end{equation}
In what follows, where convenient, we drop the subscripts $t$ and $p$ with the understanding that the discussion may be taken to apply to stability based on either typical or periodic orbits. Noting that synchronism is stable for $\xi^+>\xi>\xi^-$, and that $\xi=\gamma(1-\alpha)$, we consider the coupling network topology-dependent ratio $(1-\alpha^-)/(1-\alpha^+)$ where $\alpha^+$ ($\alpha^-$) denotes the maximum (minimum)  eigenvalue of the adjacency matrix  (not including the eigenvalue $\alpha=1$ corresponding to the eigenvector $(1,1,...,1)^T$). Recall that $(1-\alpha)\geq 0$. Since $\xi^+>\xi>\xi^-$ for stability, if
\begin{equation}
s>\frac{1-\alpha^-}{1-\alpha^+},
\end{equation}
then the system can be made stable by adjustment of the constant $\gamma$, but, if $s<{(1-\alpha^-)}/{(1-\alpha^+)}$, then it is impossible to choose a value of $\gamma$ for which $M(\xi)<0$ for all the relevant eigenvalues $\alpha$, and stability is unachievable. Figure 3(c) shows plots of $s_t$ and $s_p$ versus $\nu$ for the same parameters as used in Figs. 3(a,b). Note that, for these computations, the values of $s$ without adaption (i.e., $s_t=23.5$ and $s_p=10.5$) always exceed the corresponding values with adaption. We have also found this to be true for the generalized adaptive scheme of Sec. III B (which includes the additional adaption parameter $\phi$). However, we do not know whether this is general, or is limited to our particular example (Eq. (\ref{RC}) with $H(x)=u$, and our choices of the parameters $a$,$b$, and $c$).

To test our linear results in Fig. 3, we have also performed fully nonlinear numerical simulations for a simple network consisting of a sender system (labeled 1) connected to a receiver (labeled 2). In this case Eq. (1) becomes
\begin{subequations}\label{sr}
 \begin{align}
 \dot{x}_1(t)= & F(x_1(t)), \\
 \dot{x}_2(t)= & F(x_2(t))+\gamma \Gamma [\sigma(t) A(t) H(x_2(t))-H(x_1(t))],
 \end{align}
 \end{subequations}
 and $A(t)$ is a scalar.
\begin{figure}[h]
\centering
\includegraphics[width=0.70\textwidth]{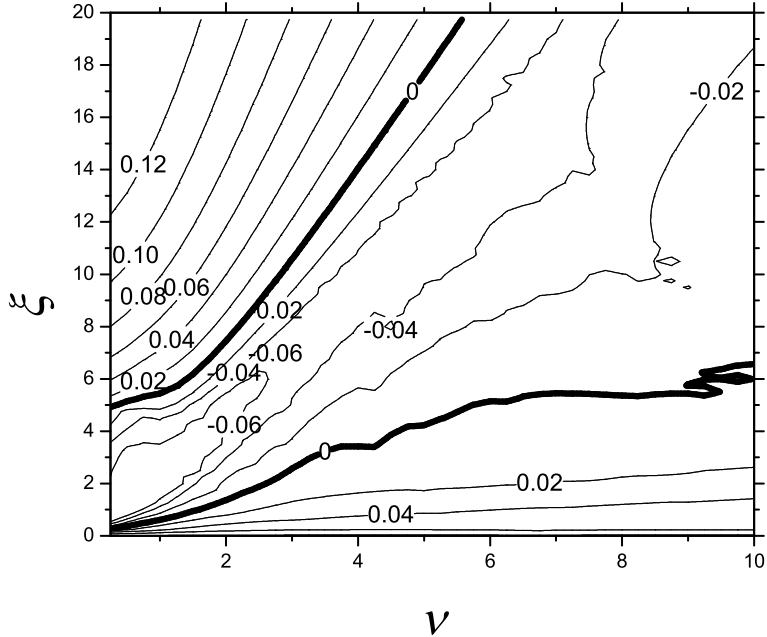}
\caption{The figure is a level curve plot in $\xi$-$\nu$ space of the values assumed by the master stability function $M$, evaluated for $x_s(t)$ being a typical chaotic orbit. The area of stability (corresponding to $M<0$) is delimited by the thick $0$-level contour line.  $F(x)$ is the R\"ossler equation (\ref{RC}),   $H(x(t))=u(t)$, and $\Gamma=[1,0,0]^T$.  
\label{DUE}}
\end{figure}
\begin{figure}[h]
\centering
\includegraphics[width=0.50\textwidth]{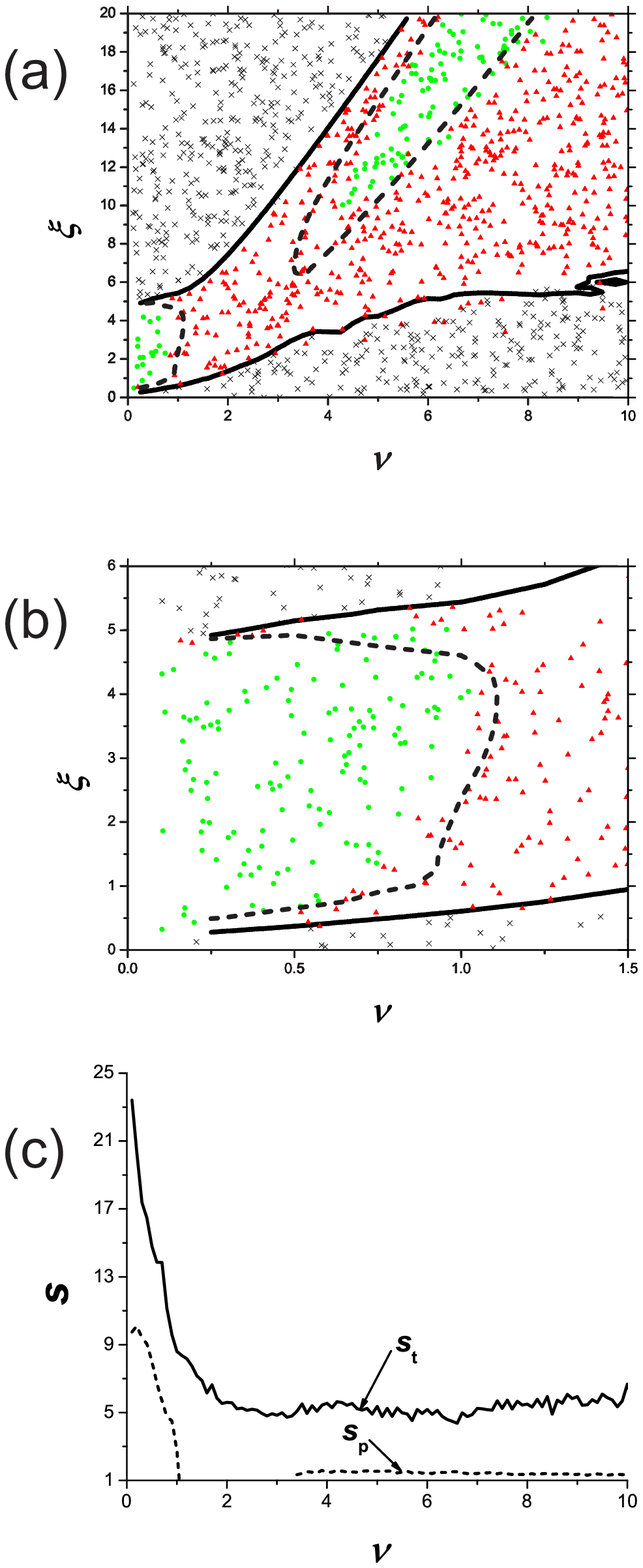}
\caption{In plot (a), thick solid curves (thick dashed curves) bound the area in which the master stability function $M(\xi,\nu)$ is negative for $x_s(t)$ corresponding to a typical chaotic orbit in the R\"ossler attractor (for $x_s(t)$ corresponding to the maximally unstable periodic orbit embedded in the attractor of period up to four), 
$F(x)$ is the R\"ossler equation (\ref{RC}),   $H(x(t))=u(t)$, and $\Gamma=[1,0,0]^T$.
Each data point shown in the figure is the result of a simulation involving a sender (maestro) system connected to a receiver, where the receiver state was initialized by a displacement of $10^{-8}$ from the sender state. A step-size of $10^{-4}$ was used for a run time of $10^5$ time units. If, in that time span, the synchronization error $E$ never converged to $0$ and, at some point, exceeded $0.1$, the run was considered to be unstable (corresponding to an $\times$ in the figure). If $E$ converged to $0$, a $1 \%$ mismatch in the R\"ossler parameter $a$ was introduced to the receiver, and the run was repeated with an initial separation of $0$. Then, if $E$ ever exceeded $0.1$, the run was considered to be bubbling (corresponding to a green circle in the figure), otherwise the run was considered to be stable (corresponding to a red triangle in the figure). Plot (b) is a blow up of the lower left corner of plot (a). Plot (c) shows the synchronizability ratios $s_t$ (solid curve) and $s_p$ (dashed curve) versus $\nu$. The missing data points for the dashed curve are a result
of the low period orbits not having a range of stability for those values of
$\nu$. We found that the synchronizability ratios for the
nonadaptive case are equal to those in the limit $\nu \rightarrow 0$. \label{DUEbis}}
\end{figure}
Each data point shown in Figs. 3(a,b) corresponds to a run, 
where the sender was given a random initial condition and random values for $\nu$ and $\xi$ were chosen in the plotted range. After waiting sufficient time to ensure that the sender state is essentially on the attractor, the $u$-variable of the receiver state was initialized by a displacement of $10^{-8}$ from the $u$-variable of the sender state.  A step-size of $10^{-4}$ was used for a run time of $10^5$ time units, over which we recorded the normalized synchronization error,
 \begin{equation}
E(t)=\frac{|u_1(t)-u_2(t)|}{<(u_{s} - <u_{s}>)^2>^{1/2}}, \label{E}
\end{equation}
where  $<...>$ indicates a time average and the subscript $s$ denotes evolution on the synchronous state (i.e., using dynamics from Eq. (\ref{xs})).
If, in that time span, $E$ never converged to $0$ and, at some point, exceeded $0.1$, the run was considered to be unstable (corresponding to an $\times$ in the figure). If $E$ converged to $0$, a $1 \%$ mismatch in the R\"ossler parameter $a$ was introduced to the receiver, and the run of duration $10^5$ time units was repeated with an initial separation of $0$. If, at any time during the run, $E$ ever exceeded $0.1$, the run was considered to be bubbling (corresponding to a green circle in the figure), otherwise the run was considered to be stable (corresponding to a red triangle in the figure). We see that the master stability computations of the high-quality stable, bubbling, and unstable regions (the solid and dashed lines) correspond well with these results.
We also did a sampling of points up to period 5 and did not find that this altered our results. From Fig. 3(a), we observe the presence of a few green circles (i.e, bubbling) within the high-quality synchronization area, delimited by the dashed line. In reference to this observation, we note that (i) for the case in which a small parameter mismatch is present, the synchronization error is expected to vary smoothly with parameter variation, and there is no sharp transition from the stable to the bubbling regime; and (ii) our computations show that close to the dashed line, the master stability function associated with the most unstable invariant set embedded in the attractor is rather small.
Facts (i) and (ii) explain our difficulty in using our nonlinear computations to clearly separate the bubbling from the stable regions about the dashed line in Fig. 3(a).
An important point concerning Figs. 3(a,b) is
that the area associated with bubbling in Fig. 3(a) is rather substantial. This observation would become particularly important in experimental realizations of adaptive synchronization, since small mismatches in the parameters and noise cannot be   avoided in experiments.

Figure 4 shows a sample plot of the normalized synchronization error $E(t)$, versus $t$. We implemented our adaptive strategy with values of $\nu=2.5$ and $\gamma=5$, corresponding to the bubbling regime (see Fig. 3), $A(t)=1$, and the receiver has $0.1\%$ mismatch in the parameter $a$. The two insets are zooms showing phase-space projections in the plane $(u_2,v_2)$, over two different time intervals. Inset (b) corresponds to a range of time between bursts ($E(t)< 5 \times 10^{-2}$) and shows that during this time the orbit is essentially that of a typical chaotic orbit. Inset (a) shows the orbit trajectory for a range of time during which a burst is growing. It is seen from inset (b) that during the time range of the growing burst the orbit closely follows a period 4 orbit embedded in the attractor. The burst is evidently caused by the instability  of this period 4 orbit to perturbations that are transverse to the synchronization manifold.
\begin{figure}
\centering
\includegraphics[width=0.70\textwidth]{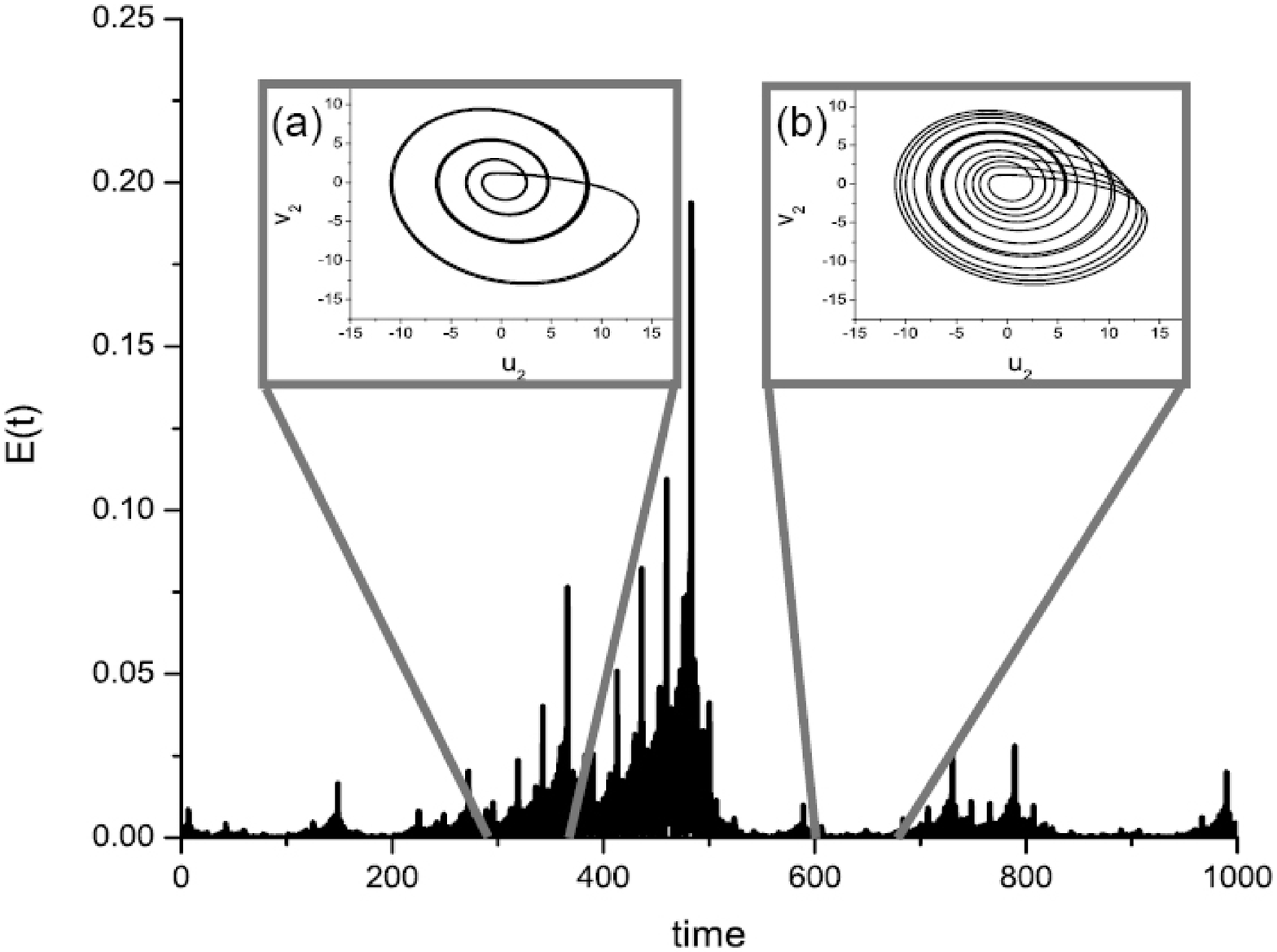}
\caption{The figure shows the synchronization error $E(t)$ versus $t$ for a simple network consisting of a sender connected to a receiver (Eqs. (\ref{sr})), $F(x)$ is the R\"ossler equation (\ref{RC}),   $H(x(t))=u(t)$, $\Gamma=[1,0,0]^T$,  $\gamma=5$, $\nu=2.5$, $A(t)=1$, $dt=10^{-3}$. The receiver has a $0.1\%$ mismatch in the parameter $a$. The two insets are zooms showing phase-space projections in the plane $(u_2,v_2)$, over two different time intervals. Inset (b) corresponds to a typical chaotic orbit for which the synchronization error is small, i.e., $E(t)<5 \times 10^{-2}$, while inset (a) corresponds to an unstable period 4 periodic orbit embedded in the attractor, for which $E(t)$ is eventually large (i.e., a burst occurs).
\label{QUATTRO}}
\end{figure}

We have also performed numerical master stability computations for our generalized adaptive strategy, presented in Sec. IIIb. This is shown in Fig. 5, where the $\xi^+(\phi)$ and $\xi^-(\phi)$ curves, corresponding respectively to the largest (smallest) values of $\xi$ for which $M(\xi,\phi)>0$ ($M(\xi,\phi)<0$), are plotted versus $\phi$ for three different values of $\nu=[0.1,2.0,6.0]$ for typical chaotic orbits. For small $\nu$ (e.g., $\nu=0.1$ in the figure), the range of stability $[\xi^-,\xi^+]$ is almost independent of $\phi$, while for larger values of $\nu$ the choice of $\phi$ can significantly affect the $\xi$-range of stability. As expected, at $\phi=2$, $\xi^+(\phi)$ and $\xi^-(\phi)$ are independent of $\nu$.
\begin{figure}[h]
\centering
\includegraphics[width=0.70\textwidth]{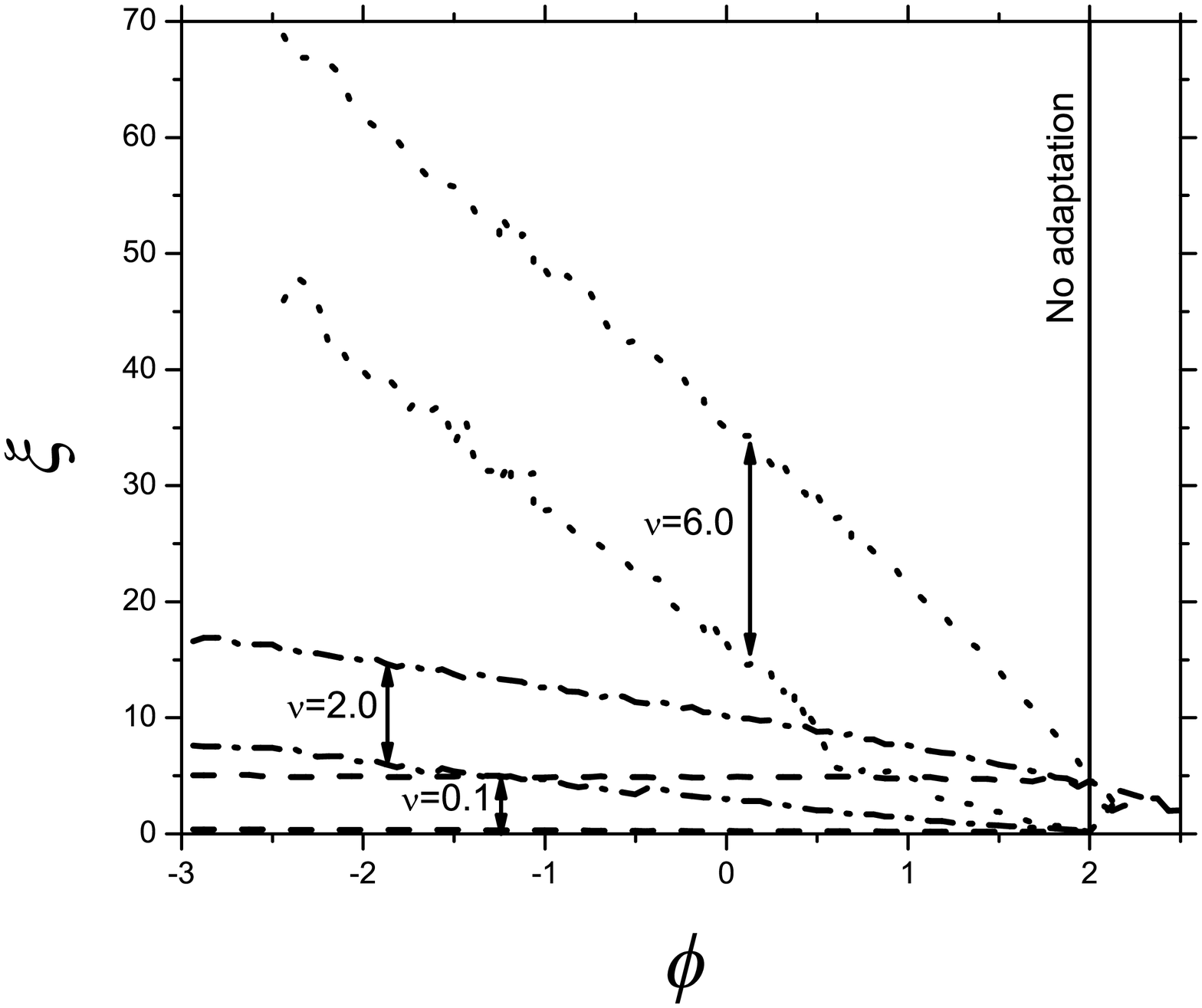}
\caption{The plot shows the area in the parameter space ($\phi,\xi$) in which $M(\xi,\phi)$ obtained from (\ref{INDA}) is negative, for three different values of $\nu=[0.1,2.0,6.0]$; $F(x)$ is the R\"ossler equation (\ref{RC}),  $H(x(t))=u(t)$, and $\Gamma=[1,0,0]^T$. The stability areas are upper and lower bounded by the $\xi^+$ curve and the $\xi^{-}$ curve, plotted as function of $\phi$. As the figure shows, at $\phi=2$, $\xi^+$ and $\xi^-$ are independent of $\nu$, corresponding to the case of no-adaptation. 
\label{CINQUE}}
\end{figure}

Finally, we investigated whether, for our coupled systems with adaptation, bubbling can be caused by a slow drift in the coupling strength.
For this purpose we now take the parameter $A(t)$ in Eq. (\ref{sr}) to have a slow time drift,
\begin{equation}
A(t)=1+0.2 \sin(2 \pi \times 10^{-3} t).
\end{equation}
We implemented our adaptive strategy with values of $\nu=1$ and $\gamma=2$, corresponding to the bubbling regime (see Fig. 3).
 For most of the time there is good synchronization between the sender and the receiver, but we also observed the intermittent occurrence of short, intense desynchronization bursts. Figure 6 shows the synchronization error $E(t)$ versus $t$. Note that in the absence of parameter drift ($A$ constant), the synchronization error would eventually become zero. This simulation shows that, similarly to the previously reported burst-inducing effect of small parameter mismatch or noise, drift also promotes the continuous intermittent occurrence of bursting. 

\begin{figure}[h]
\centering
\includegraphics[width=0.70\textwidth]{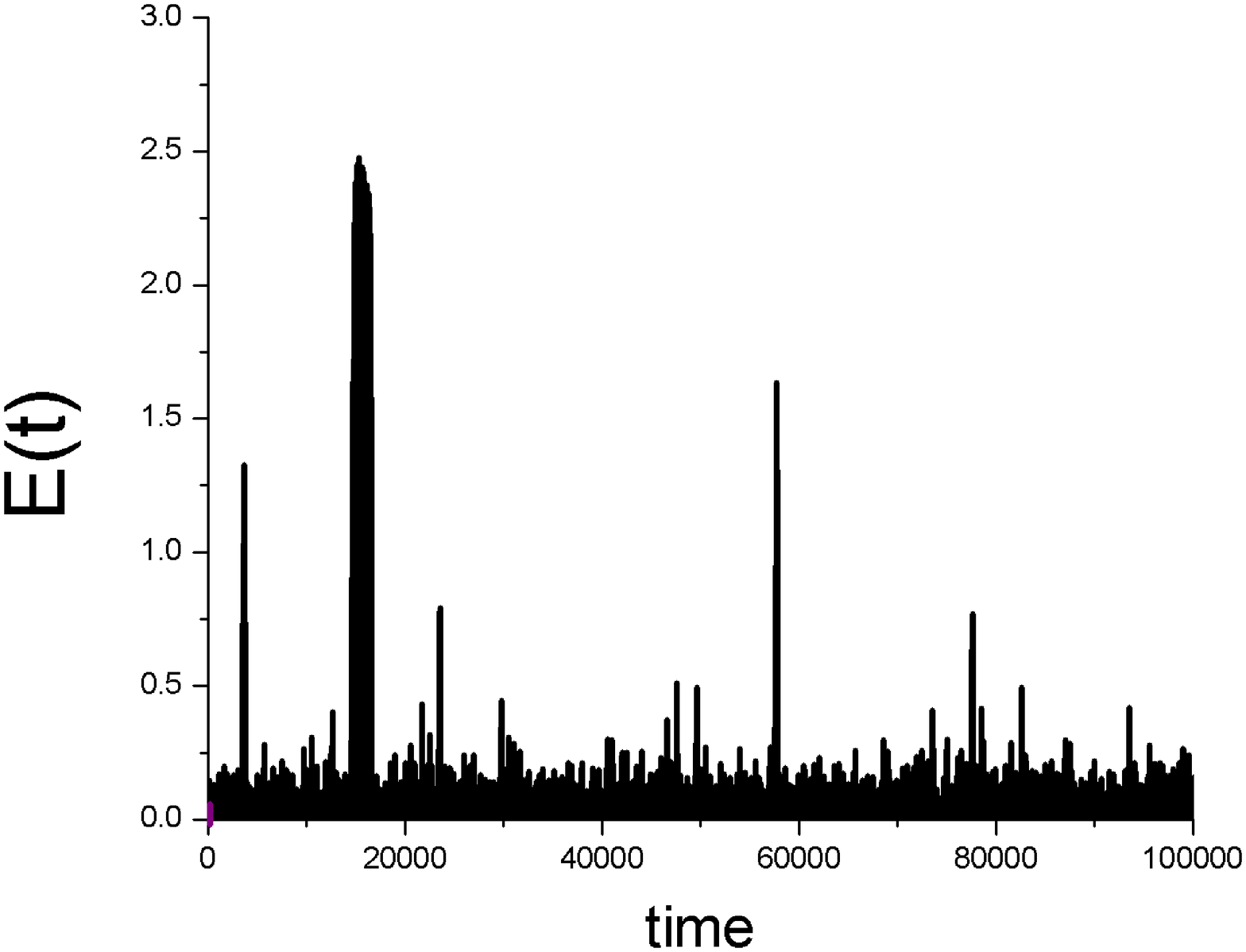}
\caption{The figure is a plot of the synchronization error $E(t)$ (defined in Eq. (\ref{E})) versus $t$ for a simple network consisting of a sender connected to a receiver (Eqs. (\ref{sr})), $F(x)$ is the R\"ossler equation (\ref{RC}),   $H(x(t))=u(t)$,  $\Gamma=[1,0,0]^T$,   $\nu=1$, $\gamma=2$, $A(t)=1+0.2 \sin(2 \pi \times 10^{-3} t)$, $dt=10^{-3}$. As can be seen, the dynamics of $E(t)$ exhibits intermittent
bursting.} 
\label{SEI}
\end{figure}

\section{Conclusion}

This paper is concerned with the study of stability of adaptive synchronization of chaos in coupled complex networks (e.g., sensor networks). As an example addressing this issue, we consider a recently proposed adaptive scheme for maintaining
 synchronization in the presence of {a priori} unknown slow temporal drift in the couplings \cite{SOTT}. In contrast with previous approaches (e.g., \cite{rr3,rr4,rr6,rr8}), based on system specific use of the  Lyapunov function technique, we present a master stability analysis which predicts the exact ranges of stability for the synchronized state. We observe that the stable range of synchronism can be sensitively dependent on the adaption parameters. Moreover, we are able to predict the onset of bubbling, which occurs when the synchronized state is stable for typical chaotic orbits but is unstable for certain unstable periodic orbits within the synchronized chaotic attractor. We define stability to be \emph{high quality} when the synchronized state is stable with respect to all the orbits embedded in the attractor and numerically find the regions of `high quality stability' for a given system of interest.
 We also found  that, for our coupled systems with adaptation, bubbling can be caused by a slow drift in the coupling strength, in addition to small noise and small mismatch in $F$. We emphasize that, since parameter mismatch, noise and drift are ubiquitous in experimental situations, and since (e.g., Fig. 3(a)) bubbling can occupy substantial regions of parameter space, consideration of bubbling can be expected to be essential for determining the practical feasibility of chaos synchronization applications.

We thank Anurag. V. Setty and Bhargava Ravoori for the enlightening discussions.

This work was supported by ONR grant N00014-07-1-0734.

\section*{APPENDIX I: Stability of the generalized adaptive strategy}

We note that the function $Q([{p_i(t) r_i(t)}]/[{q_i(t) H(x_i(t))]})$ in Eq. (\ref{quattrobis}), when evaluated about  (\ref{cinque}), is equal to one.
Then, by linearizing Eqs. (\ref{net}), (\ref{quattroa}), and (\ref{quattrobis}) about (\ref{cinque}), we obtain,
\begin{subequations}\label{A1}
\begin{align}
\delta \dot{x}_i=DF^s \delta x_i+ \gamma \Gamma \biggl\{ DH^s  \biggl[k_i^{-1} \sum_j A_{ij} \delta x_j - \delta x_i \biggl]+ \frac{H^s}{k_i^2 <(H^s)^2>_{\nu}}  \epsilon_i \biggl\}, \quad i=1,...,N, \label{A1a}\\
 \dot{\epsilon}_i=-\nu \epsilon_i + (\phi-1) \frac{(H^s)^2}{<(H^s)^2>_{\nu}} \epsilon_i +(\phi-2)  H^s DH^s \biggl[k_i \sum_j A_{ij} \delta x_j- k_i^2 \delta x_i \biggl], \quad i=1,...,N. \label{A2a}
\end{align}
\end{subequations}
As in our derivation of Eqs. (\ref{IND}), we again set $\delta x_i=c_i \bar{x}(t)$, where $c_i$ is a constant scalar that depends on $i$ and $\bar{x}(t)$ is a vector that depends on time but not on $i$. Equations (\ref{A1}), then become
\begin{subequations}\label{A2}
\begin{align}
\dot{\bar{x}}=DF^s  \bar{x}+ \gamma \Gamma \biggl[  \frac{\sum_j A_{ij}  c_j }{k_i c_i} - 1 \biggl] DH^s \bar{x}+ \frac{ \gamma \Gamma H^s}{ c_i k_i^2 <(H^s)^2>_{\nu}}  \epsilon_i, \quad i=1,...,N,  \label{A2a} \\
 \dot{\epsilon}_i=-\nu \epsilon_i + (\phi-1) \frac{(H^s)^2}{<(H^s)^2>_{\nu}} \epsilon_i +(\phi-2) H^s DH^s \biggl[k_i \sum_j A_{ij} c_j- k_i^2 c_i \biggl] \bar{x}, \quad i=1,...,N.\label{A2b}
\end{align}
\end{subequations}
To make  Eqs. (\ref{A2}) independent of $i$, we again consider $\beta'(t)=\epsilon_i(t)/[c_i {k_i} (\alpha-1)]$ and take $\alpha$ to be the eigenvalues of ${\bf{A'}}=\{A'_{ij}\}=\{{k_i}^{-1} A_{ij}\}$, resulting in Eqs. (\ref{INDA}).

\section*{APPENDIX II: Determination of unstable periodic orbits}
\par To account for the phenomenon of bubbling, it is necessary to look not just at typical (that is, chaotic) orbits of the uncoupled oscillator, but the periodic orbits embedded in the chaotic attractor as well. As there are a (countably) infinite number of such orbits, it is impossible to account for them all. However, as shown by Hunt and Ott, the optimal periodic orbits of maximal transverse instability tend to be those of low period \cite{bub2}. Thus, for our analysis, it was found to be sufficient to consider only those orbits with a period less than some appropriately chosen limit.
\par To find these low-period orbits for the R\"ossler attractor, we initialized an uncoupled oscillator with random initial conditions, waited for it to settle onto the attractor, then recorded its orbits for some suitable length of time at high temporal precision. We then noted each piercing of the surface of section $u=0$ in the positive-$u$ direction ($\dot{u}>0$). To a high degree of approximation, the $(v,w)$ coordinates of these points were found to lie a curve, thus suggesting that it is possible to reduce the three-dimensional flow to a one-dimensional map. We then plotted $v(i+n)$ vs. $v(i)$; that is, the $v$ coordinate of the $(i+n)^\text{th}$ piercing versus the $v$ coordinate of the $i^\text{th}$. Each intersection of this curve with the line $v(i+n)=v(i)$ represents the $v$ coordinate of an initial condition for an orbit that starts on the surface of section and returns to its original position after $n$ piercing of the surface of section. With two coordinates (namely $u$ and $v$) known, all that remains is to find the value of $w$ such that $(0,v,w)$ lies on the attractor.
\par Of course, for $n>1$, many of these intersections will be redundant, as every period $n$ orbit pierces the surface of section $n$ times, thus producing $n$ intersections on the curve. In addition, each curve will have intersections corresponding to orbits of any period that is a factor of $n$. As an example, consider the curve $v(i+4)$ vs. $v(i)$. The R\"ossler system used in this paper has three Period 4 orbits, one Period 2 orbit and one Period 1 orbit. Thus, the number of times $v(i+4)$ vs. $v(i)$ will intersect $v(i+4)=v(i)$  is $3\times 4+1\times 2+1\times1=15$.

As these orbits are inherently unstable, error accumulated through numerical
integration can result in a trajectory leaving the periodic orbit after only
a small number of periods. Thus, for the long term computation of Lyapunov exponents to obtain the master stability function, it is advisable to compute the
trajectory for only a single period, then return the oscillator to its
initial position $(u,v,w)$, and repeat as often as needed.

\end{document}